# [Invited Talk] Building a Disaster-resilient Storage Layer for Next Generation Networks: The Role of Redundancy

Vero ESTRADA-GALINANES†, Racin NYGAARD†, Viktor TRON‡, Rodrigo SARAMAGO†, Leander JEHL†, and Hein MELING†

†Faculty of Science and Technology, University of Stavanger, Norway
‡Swarm, Ethereum

E-mail: †{veronica.estrada,racin.nygaard,rodrigo.saramago,leander.jehl,hein.meling} @uis.no,
‡viktor@ethereum.org

**Abstract** Blockchain is the driving force behind a myriad of decentralized applications (dapps) that promise to transform the Internet. The next generation Internet, or web3, introduces a "universal state layer" to store data in p2p networks. Swarm, a native layer of the Ethereum web3 stack, aims at providing redundant storage for dapp code, data, as well as, blockchain and state data. Based on a diploma verification dapp use case, we share insights on the role of redundancy strategies in designing a reliable storage layer. Our proof-of-concept improves Swarm's resilience to failures by balancing repairs and storage, with a slightly added latency.

**Keywords** redundancy, fault-tolerance, decentralized networks, p2p networks, alpha entanglement codes, erasure codes

## 1. Introduction

Sharing information via a peer-to-peer (p2p) network is at the core of the Internet. The most popular Internet service to share information is the web, a universal publishing platform. If we delve into the foundations of the Internet, the concepts of *redundancy* and *decentralization* have a pivotal role in the design of robust distributed communication networks for a wide range of users with different requirements [1]. Centralization was discouraged since a central point is an obvious vulnerable control point. In contrast, the last decade has seen a major increase in web centralization. In addition, web content became more susceptible to privacy threats and permanent data loss.

Blockchain, together with advanced cryptography, and token economies are seen as promising technologies to build solutions to the problems mentioned above and transform the Internet. A public blockchain is an immutable, distributed ledger built and maintained within a p2p network. The ledger is a global tamper-proof data structure collectively maintained by peers without the need for trust. Peers use a consensus algorithm to consistently add new records to the ledger. Valid records are included in the blockchain (on-chain data) and replicated across all peers. This mechanism makes on-chain storage very expensive. At the time of writing, we estimate that the operation to store 1 MB of data on the Ethereum blockchain costs $1292.95 USD (6.5536 ETH). In addition, this model has hidden costs for participating peers, due to the high degree of replication: every full node must download the whole blockchain, and future participant nodes must do the same without being rewarded. Thus, this model does not take into account the costs of long-term storage.

Solutions for off-chain, decentralized storage are under intense development, e.g., IPFS[1], Filecoin[2], StorJ[3] and Swarm[4], and more services continue to emerge. In general terms, these systems will cover a wide range of requirements to build a more decentralized storage substrate for the Internet. These systems can provide cloud hosting and content distribution, while on-chain data may contain aggregates and/or pointers to off-chain data. Token economies will ease the payment of incentives to storage nodes that provide a service in accordance with contract terms. Advanced cryptography can be used to protect privacy and to audit storage nodes with proof of storage challenges. Finally, redundancy mechanisms like replication or erasure codes can be used to increase file availability and to avoid data loss due to misbehaving nodes or other types of failures and attacks. But, an open question remains, **will it be possible to retrieve a document after a decade or more with the redundancy mechanisms put in place in decentralized storage?**

In the near future, digitization will transform daily operations and paper-based documents may disappear. Thus, our society needs systems that can guarantee long-term storage with an affordable cost. Furthermore, our identities will depend more on digital content. Digital-born documents may be used to claim facts. In this scenario, decentralized storage can provide a common storage layer for exchanging verified documents between governments, institutions as well as individuals. Then, a second question arises: **is it possible to build a trustworthy distributed system to verify documents or identity claims using untrusted peers?**

To address this question, in BBChain (bbchain.no) we are designing a global system for storing and computing over degree certificates, or *diplomas*, with the following goals: (i) prevent loss of, and fraud with, diplomas and (ii) transparency and fairness in the admissions process. In doing so, we are rethinking the time-consuming paper-based processes.

In such a system for managing diplomas, public verifiability is desired. That is, a higher education institution should be able to verify the diploma of a student issued by another institution. Similarly, an employer may wish to verify the diploma of a student applying for a job. To this end, BBChain will combine biometrics and blockchains to manage digital identities and to prove the integrity of diplomas. However, due to the high degree of replication and often costly consensus mechanism [2], a blockchain would not be sustainable for long-term storage of diplomas. Instead we rely on off-chain storage for diplomas.

---

[1] https://ipfs.io/
[2] https://filecoin.io/
[3] https://storj.io/
[4] https://swarm.ethereum.org/

Our diploma application and its storage needs are characterized by the number of students and the exams they pass. As such, we expect a substantial annual data growth, yet linear in the number of active students. Diplomas are expected to be stored for the life-time of the owner. Writes to the storage will be dominated by the activities of active students, whereas we expect that diplomas will have a high mean time between reads, as individual diplomas are only needed every so often. Nevertheless, diplomas must be readily available to ensure timely verification, i.e., within a few seconds, and as such we cannot leverage cold storage. To satisfy these requirements, BBChain will store diploma integrity information on-chain, whereas the diploma documents are stored off-chain.

The contribution of this paper is twofold. First, we investigate how the BBChain system can be realized in a decentralized ecosystem. Second, we show how erasure codes, and especially alpha entanglements can enhance distributed storage infrastructure to build a disaster-resilient decentralized storage layer.

Towards the first goal, we report on our initial endeavour to implement this system using smart contracts, and Swarm [3] as storage and content delivery layer. We use smart contracts on the Ethereum blockchain to implement business logic, such as student enrollment and issuing and verification of credentials, e.g., approved courses, which may be combined into a diploma at the end of the educational cycle. We provide the big picture of the BBChain system and the general ideas behind its data model to show how a complex solution can be engineered inside the Ethereum ecosystem. We explore how to take advantage of Swarm features to build an economically viable solution.

Off-chain p2p-based storage solutions have many challenges to overcome while working towards the provision of a highly reliable service. For instance, the current version of Swarm only caches objects during uploads and downloads. Erasure codes are planned for future versions but tamper-proof documents and zero data loss require careful thinking about redundancy. We built a prototype that implements alpha entanglement codes on the client-side. In this way, we are able to control the redundancy levels aligned with our application.

Our evaluation shows that, the prototype was able to successfully retrieve the file with only slightly added latency when there are failures present in the network.

## 2. Background and Motivation

We use smart contracts and Swarm storage which are both part of the Ethereum ecosystem. Ethereum is a global, open-source platform to develop decentralized applications (dapps). The platform is supported by more than 10k nodes and a large community of developers. Inside this ecosystem, the vision of a decentralized web, or a serverless Internet, takes the name of *web3*. Decentralization is considered the next big step for the web by Internet pioneers [4]. The web3 promises that participants have: 1) the possibility to create native economic value, and 2) the possibility to transfer that value to any other participant.

The state layer (Layer 1[5]) is critical for the web3 stack as it preserves what happens in the layers below. This is possible thanks to blockchain infrastructures, i.e., state is hold and transferred natively by the blockchain. However, systems like Swarm or IPFS also build an important part of the Layer 1 infrastructure, since they allow storing and distributing large or diverse content, which is technically, or economically infeasible with on-chain data.

---

[5] https://wiki.web3.foundation

Swarm is still under development and at this stage it cannot guarantee long-term retrieval. However, its goal is to provide an adequate degree of decentralization and redundancy for the Ethereum's public record, including blockchain data, as well as, dapp's code and data.

Our analysis and evaluation show that alpha entanglements are not only viable for the BBChain system, but can help to build a disaster-resilient decentralized storage layer for the web3.

## 3. The Role of Redundancy

Redundancy plays a major role in distributed storage systems to improve consistency, fault-tolerance, and/or load balance. In essence, the dissemination of information in a network increases the redundancy at the system level. The opposite would be to increase the redundancy at the peer level, e.g., by adding redundant disk arrays, but from the system perspective that creates centralization and single point of failure.

In this section, we show how redundancy affects the dissemination of information. We start by examining two perennial mantras for redundancy:

**Redundancy Mantra #1: The More, the Better**

The more, the better is the mantra for replication. Perhaps, the system "Lots of Copies Keep Stuff Safe (LOCKSS)" [5] is the quintessence of this mantra. LOCKSS is a p2p system with a long trajectory, built to facilitate collaboration between libraries to guarantee permanent web publishing.

A system that keeps $n$ replicas for each file can tolerate $n-1$ failures. It is also well-known that $n$ replicas can be used to build a reliable distributed system that tolerates $f = (n-1)/3$ Byzantine failures if the nodes who keep the replicas reach an agreement in an election protocol [6]. However, the authors of LOCKSS proposed that the number of replicas should be $n \gg 3f + 1$ to relax the voting procedure into opinion polls where peers decide over version consistency [7].

Bitcoin embraces this mantra by building a fully decentralized global currency system based on a replicated layer, the blockchain. The Bitcoin network disseminates transactions and blocks to all peers. The blockchain is built collectively based on a consistent view of the propagated information [8].

**Redundancy Mantra #2: The Less, the Better**

The less, the better is the mantra for Maximum Distance Separable (MDS) erasure codes. MDS erasure codes aim to increase fault tolerance with minimal space requirements. If a block has size $B$, the encoding method splits the object in $m$ fragments with size $B/m$ and generates $k$ equal-size fragments. Then, it distributes the total $n=m+k$ fragments in $n$ peers. Protocols that tolerate Byzantine failures such as the $m+3f$ protocol need to distribute fragments in more peers [9]. MDS erasure codes are acclaimed for being storage efficient, an erasure coded system tolerates $k$ failures with a redundancy factor of $(m+k)/m$, while the replicated alternative uses $k+1$. Object reconstruction is possible when $m$-out-of-$n$ fragments are available. Tahoe Lafs [10] is an emblematic example of an erasure coded p2p system.

The use of coding in storage systems brings associated a number of limitations, which are highlighted here. The repair mechanism depends on the availability of $m$ peers (one missing fragment is reconstructed with $m$ fragments) and it reads $m$ times more information (repairing all the fragments stored in a failed peer with storage capacity *C* bytes requires reading *m x C* bytes from other peers). Finally, maximum capacity is achieved when *n* is large but in storage systems *n* is typically a small number [11].

The literature is abundant with comparison studies concerning these two mantras, finding trade-offs between coding and replication [12-16]. A careful reading reveals that implementing redundancy in a p2p system is less than trivial and that a storage system may find compromises with a hybrid strategy that combines both mantras. However, a hybrid strategy requires a more complex architecture and in many cases removes the savings of erasure codes. Substantial efforts were dedicated to reduce the repair costs using network codes [17], but the alternatives are still limited by the underneath code. Furthermore, the implementations bring associated other complexities and, although some deterministic decentralized algorithms have been proposed [18], we are not aware of any practical implementation in storage systems.

If a storage system fails to do proper maintenance, the redundancy of each file decays until some files become irrecoverable. Poor maintenance is an insidious problem that is more difficult to solve in decentralized p2p systems. The whole system will suffer if files become irrecoverable. We hypothesize that peers will be prompted to collaborate if repairs are simple and have few requirements. Thus, our own mantra is the following:

**Mantra #3: Balancing repair and space requirements**

Alpha entanglement (AE) codes [19] try to balance decoding complexity and space requirements. These codes are xor-based, non-MDS codes. An entangled storage system can provide high fault-tolerance with little encoding and decoding complexity. They stand out for the low cost of repairing single-failures, which can be used recursively or in parallel in cases of multiple failures. In any configuration, a single-failure is repaired by a simple bitwise xor of two blocks. The parameter α indicates the number of redundant blocks generated for each information block. That is to say, a system that has four replicas has the same storage overhead as with α=3. The redundant blocks are bitwise xor-ed with other information blocks to generate new redundant blocks in a way that redundancy keeps propagating across all storage locations.

**Dissemination of Information**

Figure 1 illustrates key aspects of these redundancy mantras assuming an information block of size *B* bytes that may be split into smaller fragments. The chart shows how many peers are required to repair a single failure and how many peers store fragments, which are relevant for the repair of our block. Additionally, the bubble size shows the replication factor, i.e., the space requirements of the different schemes. At the bottom of the bubble chart it is indicated how data blocks are sharded (split) and how many redundant blocks are created.

The replication scheme *R*=4 tolerates three failures with a system space requirement of 4*B* bytes. Data and redundant blocks are spread in four peers. Replication stands for the lowest peer availability requirement, if a block is not available only one peer is required for repairs. With mantra #1, the system needs to add *B* bytes for each peer that can potentially contribute in repairs.

The scheme 4-out-of-7 splits the original block into four fragments, and three redundant fragments are encoded to tolerate three failures. This scheme has less system space requirements, only 7/4*B* bytes, but the minimum peer availability requirement increased four times. The 8-out-of-14 scheme shows how to double fault-tolerance with the same space requirements by halving the size of the fragments. This scheme spreads data and redundancy more widely (14 peers) and each involved peer only stores *B*/8 bytes. However, doing so doubles the number of peers that need to be available when one fragment is missing. The next scheme 4-out-of-16 expands the number of redundant fragments but data fragments are not additionally sharded. Since, more redundant fragments are encoded, the scheme tolerates twelve failures. This strategy results in a higher space requirement (equivalent to *R*=4) but the peer availability requirement does not increases. Fragments are spread into sixteen peers.

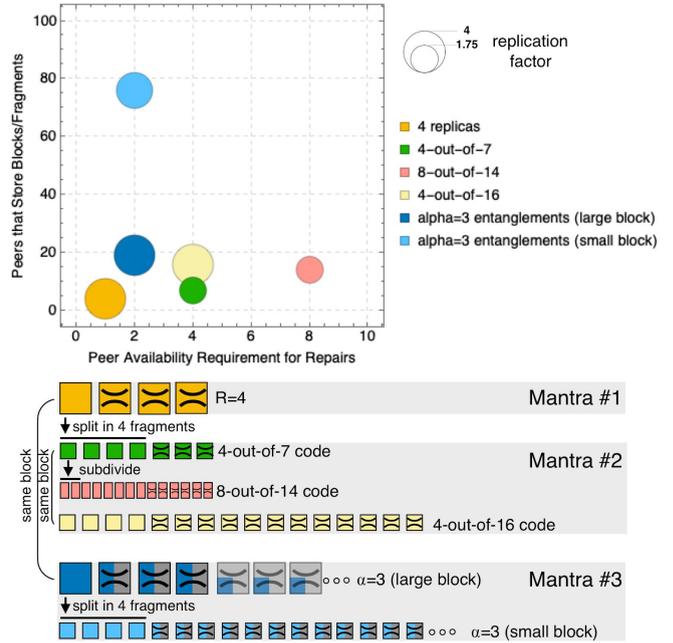

Figure 1: Redundancy mantras and dissemination of information.

The last two cases correspond to α=3 entanglements with system space requirements equivalent to *R*=4. The second case uses sharding to protect the same information with smaller blocks (*B*/4 bytes each). In these schemes, each data block generates three (α) redundant blocks. These blocks do not only contain redundant information for the given data block, but also for other data items (shown in gray). Similarly, redundant information of our block is spread throughout the system, as indicated by the additional blocks.

Important aspects to highlight: a) no matter the block size, repairs have a fixed peer availability requirement of two peers only (this property remains valid with any α value), and b) the scheme spreads data and redundancy to a larger number of peers. Due to the block dependencies created by the encoding algorithm, other blocks in the system (entangled blocks) can help to recursively repair the information blocks via their direct associated redundant blocks. Only some block combinations repair a particular block, but previous work showed that entanglements outperform *m*-out-of-*n* in realistic scenarios (small *n*) using the same storage capacity [19]. Point b) is reflected in Figure 1, as a large number of peers that store fragments, relevant for the repair of our data items. For each set of data block and its alpha redundant blocks, there are fifteen additional entangled blocks in the network that could be used to repair elements of the set (a total of nineteen peers are involved if we consider only the peers that could participate in a first or second round of repairs). The case of small blocks involves seventy six peers.

## 4. Decentralized Storage Layer

Serving files on demand and preserving the integrity of them is critical to any storage system. However, when the system is decentralized, a unique set of challenges emerges in order to provide the same guarantees. Namely, how can untrusted peers provide persistent storage. In principle, a decentralized, distributed, and immutable database can be built atop a blockchain. But this sole approach does not scale. Blockchain is a distributed data structure that follows redundancy mantra #1 but, obviously, it is economically non-viable to replicate all information in the world. Therefore, off-chain storage solutions are gaining momentum. Many options could be used in combination with an Ethereum blockchain. Swarm is deeply integrated with Ethereum via the devp2p protocol. This section describes Swarm and the rationale behind our decision to study redundancy in Swarm.

### 4.1 Why Swarm?

Swarm provides a permissionless, decentralised storage and communication infrastructure. Swarm is still in an early stage of development, but a public testnet exists and during a 4-month study NodeFinder [20] found 6,500 Swarm peers. While anyone can join this system, its main component is made up of nodes that the Ethereum Foundation runs on the AWScloud. Swarm client is an open source development written in Golang. It's latest version 0.5.0 released in September, many features are still in development.

At the present time, Swarm does not offer persistent guarantees. The system only disseminates redundancies during the syncing process. Files are uploaded to a local Swarm node (or gateway) and then synced to the network. The network only operates with chunks (the notion of files does not exist). The client splits a file in multiple chunks and computes their hashes, which are used as addresses (keys) to decide the *network store peer*s (nodes that store the chunk) and to retrieve the chunks from the network. This is achieved via the distributed preimage archive (DPA), which is similar to a distributed hash table that implements a key-value store. During syncing nodes that are part of the routing path, close to the store peer keep a copy of the chunk. Peers reclaim space when the maximum resource utilization is reached by purging least accessed chunks. The system will implement storage insurance to protect content from being deleted.

*Redundancy is of major importance for Swarm.* It should be deeply embedded in its design to provide fault-tolerance, censorship resistance, DDoS resistance, and zero downtime. But at the same time, redundancy should be easy to maintain to ensure network's economic viability.

### 4.2 Swarm Overview

**Swarm nodes (peers):** Peers are identified by the hash of the Ethereum address of the Swarm base account.
**Network:** The network relies on the Ethereum devp2p rlpx suite. RLPx is used for node discovery and routing; it is based on Kademlia [21]. Peer connections are established over devp2p.
**Distributed preimage archive (DPA):** The DPA is a distributed hash table that implements a key-value store.
**Swarm gateway:** It makes possible to interact with the network via a browser and the bzz protocol.
**Pyramid chunker:** The chunker splits data streams (or documents) into chunks as well as joins chunks into documents.
**Chunk:** The basic unit for storage and retrieval is a chunk of limited size (4K bytes). A chunk is stored in plain text, but they are optionally encrypted. The Swarm network only observes chunks.
**Addresses:** Chunks are identified with a 32 byte hash address computed with the Keccak 256 SHA3 hash function applied to the concatenation of the chunk length and the chunk payload.
**Merkle tree (bzzhash):** Swarm uses Merkle trees to represent files, where the leaves are data chunks, and the intermediate nodes are tree chunks that are used to compute the root chunk. Each tree chunk contains up to 128 address of 32 bytes each.
**Files:** The file hash points to the root chunk of a Merkle tree.
**Manifest:** It is a json file containing a list of entries, which can be used as a routing table or a key-value index with integrity guarantees. Each entry defines a content that can be retrieved through the manifest hash, making possible to create hierarchical structures by composing manifests, like a file system or a DHT.
**Bzz URL scheme:** Swarm exposes the manifest API to enable URL based addressing. The bzz protocol assumes that the content referenced in the URL is a manifest and renders the content entry whose path matches the one in the request path.

## 5. Swarm Use Case: BBChain

In this section we identify how the BBChain project can addresses the domain of verifiable credentials [22], making use of the Ethereum blockchain as a verifiable data registry and the Swarm as file storage and key-value database, to establish a tamper-resistant ordering relation of users' credentials. To accomplish this goal, BBChain focus initially on the academic credentials of the educational sector. Institutions play a role of notaries, issuing cryptographically verifiable diplomas that are chronologically bundled together through smart contract invocations, allowing the easy detection of frauds based on cryptographic primitives and providing audition mechanisms. The diplomas are stored in Swarm and only a proof of emission or revocation is stored in the blockchain. Furthermore, the students can keep track of their certificate updates using a builtin asynchronous messaging system in Swarm, namely feeds.

### 5.1 Storage Requirements for Diplomas

Academic certificates are documents considered immutable and life-long, thus any digital system that intent to store digital diplomas must guarantee the following properties: 1) fault-tolerant and long-term storage, 2) document integrity and availability, 3) public verifiability, and 4) respect user's privacy. Additionally, the system needs to be economically viable. Furthermore, due to the increasing student mobility it is desirable that a diploma issued by one country is computationally verifiable by institutions or companies in another country, even if the issuing institution doesn't exist anymore or the countries don't maintain diplomatic relations.

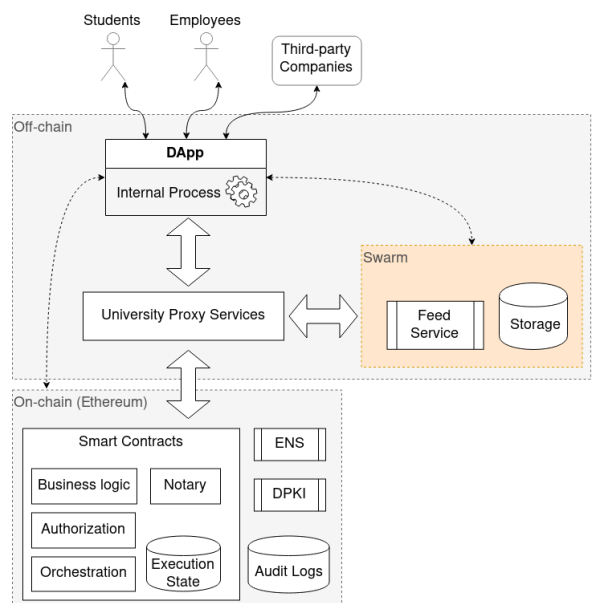

Figure 2: Overview of BBChain architecture.

### 5.2 BBChain Architecture Overview

Figure 2 shows an overview of the BBChain architecture. It illustrates the interaction of agents in an institution, e.g,. a university, with the other components of the system. The institution acts as a proxy service for the users, connecting with the Ethereum blockchain and Swarm. Preferably, the institution will maintain peers of both networks, but the client dapp can decide to connect directly to other peers (shown by the dotted arrows).

The use of smart contracts enhances the trust of the institutional procedures while reducing bureaucracy, removing the dependence on human judgment in the verification process of diplomas, and the necessity of trusted central authorities. Each contract acts as a public notary controlled by the institution. Smart contracts implement all critical business logic of the university, as well as authorization policies for services like: assigning grades or issuing diplomas. The contracts are organized hierarchically by school department or roles, giving more flexibility while keeping a history of all actions taken.

The "actions" represent some institutional procedures and are logged as events in the blockchain. Actions that require some proof of authenticity, e.g. issuing diplomas, are registered in the contract's state as certificate_proofs. The dapp creates the diplomas and stores them in Swarm following a hierarchical manifest construction.

The diplomas can be easily retrieved from Swarm based on the trusted hashes stored in the contracts.

# 6. Proof of Concept

The current version of Swarm only caches objects during uploads and downloads but does not otherwise protect against data loss. Our prototype improves file resilience in Swarm using the redundancy mantra #3, that is to say, files are encoded using alpha entanglement codes and fragments are spread over Swarm peers.

The prototype[6] adds client-side redundancy by an additional layer that entangles content, e.g., a file block with another one. Implementing entanglements on the client side allows specifying the application specific redundancy settings. This also ensures that encoding/decoding is independent of versioning and maintenance on the backend. Finally, our prototype can give useful insights and shows the feasibility of a future integration with Swarm.

In this implementation, each entangled block fits into a single Swarm chunk. Each of the newly constructed chunks and parities is then uploaded as individual files to Swarm, and the resulting manifest hash is kept in local storage for later recovery. These approach introduces overhead since Swarm creates a manifest for each block instead of only one for a file. On the other side, building the prototype atop the Swarm client is more flexible in case we need to replace the storage layer for future needs.

## 6.1 Entanglement Layer Implementation

Alpha entanglement codes, specified as AE($\alpha$,s,p) codes, serve as a mechanism to propagate redundant information in a storage system [19]. On encoding, these codes create $\alpha$ parity blocks for every data block. Parity blocks contain the bitwise exclusive-or (xor) between their data block and an existing parity block from the system, thus entangling the different parity blocks. Entanglement codes allow repairing a data-block by computing the xor between two parity blocks. For any data block $\alpha$ such pairs exist. Additionally, since parity blocks are entangled with each other, it is possible to repair a parity block by xor-ing other blocks from the system.

Based on this, we implemented two simple strategies for repair, *hierarchical* and *round robin* (see Figure 3). Both try to repair a missing data block by retrieving two parities. They differ however, if one of the parities is missing. In hierarchical repair we try to retrieve additional blocks that allow repairing the missing parity. This strategy extends recursively if additional blocks are missing. In round robin repair, we first try to retrieve the other tuples that allow repairing the data block. Only if one or two parities from all $\alpha$ tuples are missing, does round robin repair resort to repair the parities using hierarchical repair.

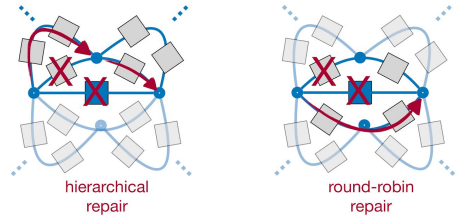

Figure 3: Strategies to repair a data block in a double failure.

## 6.2 Evaluation

We evaluate the decoding performance of our prototype by retrieving previously uploaded data from Swarm. For all our tests we run the decoding on an Intel i7 3 GHz CPU, with 8 GB of DDR3 RAM and 256GB SSD hard drive. We use the public swarm network, *swarm-gateways.net*, which is maintained by the Swarm core team. Each test is evaluated 600 times, and uses a 1 MB file, which is slightly larger than the expected size of a BBChain diploma in PDF format.

To have a baseline for comparing the effects of entanglements, we initially evaluated the performance using the standard Swarm API. Following this we entangled the same file using the configuration AE(3,5,5). Then, we ran a series of tests to evaluate the repair strategies by simulating the unavailability of data- and parity-blocks, with unavailability ranging from 0% up until 15%. Figure 4 shows the mean download times and standard mean error.

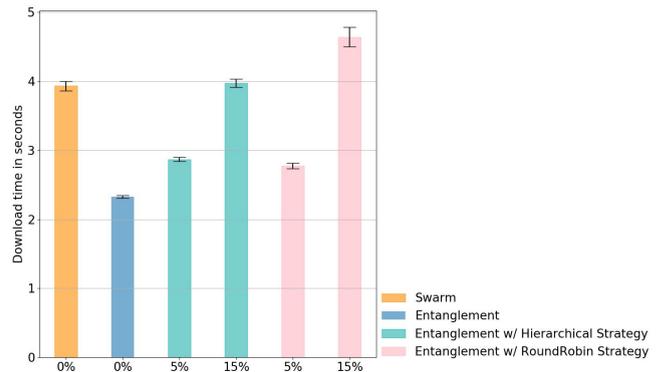

Figure 4: Downloading a 1MB file with x% of injected failures.

During our initial evaluations, we noticed that we were occasionally unable to retrieve some chunks after having recently uploaded data to Swarm. This is due to the swarm gateway load-balancing the HTTP requests. A particular request can end up at a node which is not fully synchronized yet. Therefore, we waited for up to 1 day, which seems to be a sufficient time to let all the nodes synchronize before running our experiments.

The time to download the complete file from Swarm is significantly larger than the time used by our prototype, to download data blocks in the absence of failures. We believe that this is due to the prototype implementation issues a large number of requests for data in parallel, whilst the concurrency of the Swarm baseline is controlled by the gateway.

As can be seen from the entanglement tests, we are getting the expected behavior of higher latency the larger the failure percentage is. This is due to the fact that the repair algorithm has to request and process additional parity blocks. However, also the increased latencies are reasonable compared to the Swarm latency.

Both repair strategies are aggressively tuned to sacrifice bandwidth to achieve lower latency. This means that if some parities are slow to download, additional parities are requested. Due to this, more parities than the necessary are downloaded.

---

[6] A preliminary version, written in Golang, was implemented in 2 days at the Ethereum Hackathon Madrid 2019 and won the first prize.
https://github.com/racin/HackathonMadrid_Entanglement/tree/final

When comparing the two repair strategies, we noted that the average number of parity blocks requested for each failed data block is lower in the roundrobin repair mode, whilst the minimum is higher. With 5% failure rate, we observed that the repair algorithm, on average requested 2.24 parity blocks in the hierarchical, and 2.14 in the round robin strategies. At minimum the hierarchical requested only 1.75 parity blocks for every failure, whilst in the round robin strategy 2 parity blocks were requested. From this we can observe that the hierarchical strategy has a larger variance than round robin when it comes to parity blocks requested. Interestingly, this is reversed when we look at latency, as the variance of round robin is larger in this case. With 15% failure rate, we observe the same pattern.

We believe that the repair strategies show promising results, and going forward we want to further improve upon them.

### 6.3. Discussion

Our study evaluates one of many redundancy strategies that could be used in order to increase the network resilience. In particular, we show that redundancy can be added on the client side, independent of the strategies implemented at the system level. We found that when the system experiences up to 15% data loss, the client will only experience slightly increased latency in retrieving his data. The advantage of client-side redundancy is that the user has the freedom to choose the degree of redundancy for each file. Further developments will introduce redundancy to improve: 1) uploading and downloading complete files, 2) availability of single chunks, and 3) chunk and file durability with storage insurance.

Swarm envisions a hybrid strategy to protect content that builds upon the three redundancy mantras discussed in this article. Entanglements permit to repair single blocks using 2 blocks instead of *m* in a scheme like *m-out-of-n*. This is important if the client needs to access the file partially, in which case erasure codes require prohibitive bandwidth. In addition, entanglements provide a high level of fault-tolerance. These features could be exploited in Swarm's future storage insurance layer [23]. At this layer, insurance peers will increase network resilience by adding and maintaining redundancy to insured chunks.

### 7. Conclusion

The central aim of this article is to present preliminary results of our attempts to use Swarm in the BBChain project. This contribution is relevant to increase the resilience of Ethereum's web3 storage layer.

We focus on two main BBChain's requirements: the need of long-term storage guarantees and the need for flexible features that let us combine on-chain and off-chain storage in the BBChain's business logic to verify documents. Swarm envisions to provide long-term storage and provide redundant store for apps as well as blockchain. At the moment it only adds replication during file syncing but other redundancy schemes are considered, including classic erasure codes and entanglements. Towards this goal, we show how the dissemination of information is affected with current practices of storing data redundantly. We develop a prototype that implements entanglements on the client-side, i.e., before syncing files to the network. We expect to complete this work by applying entanglements after file syncing. With respect to the BBChain smart contracts, future research will also consider a hybrid setting that combines centralized and decentralized operations.

---

[7] If you can't open any of our references it's because entanglement codes are not put into production yet